\documentstyle[preprint,aps]{revtex}
\begin{document}
\draft
\preprint{VPI-IPPAP-97-8}

\newcommand{\be}{\begin{equation}}
\newcommand{\ee}{\end{equation}}
\newcommand{\bea}{\begin{eqnarray}}
\newcommand{\eea}{\end{eqnarray}}
\newcommand{\rd}{\partial}

\title{{Vortex solutions of parity and time reversal invariant
Maxwell-Dirac-Chern-Simons gauge theory}\footnote{
Revised version (hep-th/9702124)}}
\author{{Junsoo	Shin}\footnote{Previous	address: Department of Physics,
Univ. of South Carolina, Columbia}}
\address{Department of Physics,\\
Virginia Polytechnic Institute and State University, Blacksburg, VA 24061-0435.}

\maketitle

\begin{abstract}
We construct parity and time reversal invariant Maxwell-Chern-Simons
gauge theory coupled to fermions
with adding the parity partner to the matter and the gauge fields,
which can give nontopological vortex solutions
depending on the sign of the Chern-Simons coupling constant.
\end{abstract}



Abelian gauge theories in (2+1)-dimensions with Chern-Simons terms have
been found to admit interesting classical soliton solutions [1-5].
Especially, the fermionic field theories coupled to the Chern-Simons
gauge field without Maxwell term admit vortex solutions \cite{ferm}.
Recently, it has been found that fermionic field theories coupled to
the Maxwell-Chern-Simons gauge field also admit vortex solutions \cite{shin}.
The Chern-Simons term has generally been believed to violate parity P and
time reversal T \cite{viol}.
However, it has been known that one can restore the parity and time reversal
in Chern-Simons gauge theory by adding the parity partners to the matter
and the gauge fields \cite{inva}.
It is purpose of this work to show that	such parity P and time reversal	T
invariant Maxwell-Dirac-Chern-Simons gauge theories can
give the nontopological vortex solutions with binding energies.

We consider the Lagrangian
\begin{eqnarray}
{\cal L} &=&-{1\over 4} F^{\mu\nu}_+ F_{+,\mu\nu}
-{1\over 4} F^{\mu\nu}_- F_{-,\mu\nu}
+ {\kappa\over 4} \epsilon^{\mu\nu\rho} A_{+, \mu} F_{+, \nu\rho}
- {\kappa\over 4} \epsilon^{\mu\nu\rho} A_{-, \mu} F_{-, \nu\rho} \nonumber \\
& &+ i\bar\psi \gamma^\mu \rd_\mu \psi - m\bar\psi \psi
+ eA_{+, \mu}(J^\mu_+ + lG^\mu_+) + eA_{-, \mu}(J^\mu_- - lG^\mu_-)
\label{lagrang}
\end{eqnarray}
where
$$F^{\mu\nu}_\pm = \rd^\mu A^\nu_\pm - \rd^\nu A^\mu_\pm , \qquad
J^\mu_\pm = \bar\psi \gamma^\mu P_\pm \psi , \qquad
G^\mu_\pm = \epsilon^{\mu\nu\rho} \rd_\nu J_{\pm, \rho} $$
and $\kappa$ and $\l$ are coupling constants.
The projectors $P_\pm$ onto the spin up and spin down fields are
$$P_\pm = {1\over 2} (1 \pm \rho _3 ) ,$$
where $\rho _3$ is a Dirac matrix defined by
$\rho_i \equiv \sigma_i \otimes I$.

As in Ref.\cite{shin} and \cite{topo} ,
we also introduce the topological current
$G_{\pm, \mu}$, associated with the electromagnetic current $J_{\pm, \rho}$,
which describes the induced charge and current density.
The topological current takes the form,
\be
G^i_{\pm} = \epsilon^{ij0} \rd_j J_{\pm , 0}
\label{topl}
\ee
which is related to the induced current by
\be
G^i_{ind, \pm} = \pm l \epsilon^{ij0} \rd_j J_{\pm , 0}
\label{indu}
\ee
in the system \cite{shin} \cite{topo} .
This induced current comes from the magnetic dipole moment density,
\be
\vec m_\pm = {\mu_\pm \over Q_\pm} J_{\pm, 0} \hat z
= \pm l J_{\pm, 0} \hat z
\label{dipo}
\ee
through the relation $\vec G_{ind, \pm} = \nabla \times \vec m_\pm
$\cite{shin} \cite{topo} .
Here, total charges are given by $Q_\pm = \int J_{\pm, 0} d^2 r $
and the magnetic dipole moments are $\mu_\pm \hat z$.

We choose the $\gamma$- matrices as
\be
\gamma^0 = \left( \begin{array}{cc}
\sigma_3 &0\\0 &-\sigma_3 \end{array} \right) , \quad
\gamma^1 = \left( \begin{array}{cc}
i\sigma_2 &0\\0 &i\sigma_2 \end{array} \right) , \quad
\gamma^2 = \left( \begin{array}{cc}
-i\sigma_1 &0\\0 &-i\sigma_1 \end{array} \right) .
\label{gamma}
\ee
We write $4$-component spinor field $\psi$ as $\psi =
\left( \begin{array}{c} \psi_+\\ \psi_- \end{array} \right)$
where $\psi_+$ and $\psi_-$ are two-component spinor functions.

The charge conjugation, parity and time-reversal transformation are
defined as follows\cite{ycho}:\\
1) Charge conjugation
\bea
A_{\pm , \mu}(x) \qquad &&\rightarrow \qquad -A_{\pm , \mu}(x) , \nonumber\\
\psi_\pm (x) \qquad &&\rightarrow \qquad \sigma_1 \psi^\dagger_\pm (x) .
\label {charge}
\eea
2) Parity
\bea
A_{\pm , 0}(t, x_1 , x_2 ) \qquad &&\rightarrow \qquad
A_{\mp , 0}(t, -x_1 , x_2 ) , \nonumber\\
A_{\pm , 1}(t, x_1 , x_2 ) \qquad &&\rightarrow \qquad
\mp A_{\mp , 1}(t, -x_1 , x_2 ) , \nonumber\\
A_{\pm , 2}(t, x_1 , x_2 ) \qquad &&\rightarrow \qquad
A_{\mp , 2}(t, -x_1 , x_2 ) , \nonumber\\
\psi_\pm (t, x_1 , x_2 ) \qquad &&\rightarrow \qquad
\sigma_1 \psi_\mp (t, -x_1 , x_2 ) .
\label {parity}
\eea
3) Time-reversal
\bea
A_{\pm , 0}(t, x_1, x_2 ) \qquad &&\rightarrow \qquad
A_{\mp , 0}(-t, x_1 , x_2 ) , \nonumber\\
A_{\pm , i}(t, x_1 , x_2 ) \qquad &&\rightarrow \qquad
-A_{\mp , i}(-t, x_1 , x_2 )  \quad (i = 1,2), \nonumber\\
\psi_\pm (t, x_1 , x_2 ) \qquad &&\rightarrow \qquad
\pm \sigma_1 \psi^*_\mp (-t, x_1 , x_2 ) .
\label {time}
\eea
With this definition the Lagrangian density (\ref{lagrang}) is invariant
under the parity and time-reversal transformation as well as PCT.

The equations of motion are
\be
\rd_\nu F^{\nu\mu}_\pm \pm {\kappa \over 2} \epsilon^{\mu\nu\rho}
F_{\pm, \nu\rho} = -e({J^\mu_\pm} \pm l G^\mu_\pm ) ,
\label{eon}
\ee
and
\bea
&& \gamma^\mu (i \rd_\mu + e P_+ A_{+, \mu} + e P_- A_{-, \mu}) \psi -m \psi \nonumber\\
&& - el\epsilon^{\mu\nu\rho} (\rd_\nu A_{+,\mu}) \gamma_\rho P_+ \psi
+ el\epsilon^{\mu\nu\rho} (\rd_\nu A_{-,\mu}) \gamma_\rho P_- \psi = 0 .
\label{eom}
\eea
We choose the gauge $A_{\pm, 0} = 0$ and consider the gauge field
$A_{\pm, i}$ to be static.
By taking the fermion fields $\psi_+$ and $\psi_-$ in component form,
$$
\psi_+ = \left( \begin{array}{c} \eta_+ \\ \xi_+ \end{array} \right)
e^{-iE_+ t} , \qquad
\psi_- = \left( \begin{array}{c} \eta_- \\ \xi_- \end{array} \right)
e^{-iE_- t} ,
$$
the equations of motion (\ref{eom}) can be written as
\bea
&&[ E_+ - m -el \epsilon^{ij0} \rd_j A_{+,i} ] \eta_+ = -D_{+ +} \xi_+
+ it [(i\rd_1 + \rd_2) E_+ ] \xi_+ ,\nonumber\\
&&[ - E_+ - m +el\epsilon^{ij0} \rd_j A_{+,i} ] \xi_+ = -D_{+ -} \eta_+
+ it [(-i\rd_1 + \rd_2) E_+ ] \eta_+
\label{aeo}
\eea
and
\bea
&&[ -E_- - m - el\epsilon^{ij0} \rd_j A_{-,i} ] \eta_- = -D_{- +} \xi_-
+ it [(i\rd_1 + \rd_2) E_- ] \xi_- ,\nonumber\\
&&[ E_- - m + el\epsilon^{ij0} \rd_j A_{-,i} ] \xi_- = -D_{- -} \eta_-
+ it [(-i\rd_1 + \rd_2) E_- ] \eta_-
\label{beo}
\eea
where $D_{+ \pm} = \pm i D^+_1 + D^+_2 , \quad
D_{- \pm} = \pm i D^-_1 + D^-_2$ and $D^\pm_i = \rd_i - ieA_{\pm , i}$.

If we let
\be
\psi_+ = \eta_+ \left( \begin{array}{c} 1\\0 \end{array} \right) e^{-iE_+ t},
\qquad
\psi_- = \xi_- \left( \begin{array}{c} 0\\1 \end{array} \right) e^{-iE_- t},
\label{ansaa}
\ee
the equations of motion (\ref{aeo}) and (\ref{beo}) become
\bea
&&(E_+ - m - el\epsilon^{ij0} \rd_j A_{+,i} ) \eta_+ = 0 ,\nonumber\\
&&D_{+ -} \eta_+ -it[(-i\rd_1 + \rd_2 ) E_+ ] \eta_+ = 0
\label{ase}
\eea
and
\bea
&&(E_- - m + el\epsilon^{ij0} \rd_j A_{-,i} ) \xi_- = 0 ,\nonumber\\
&&D_{- +} \xi_- -it[(i\rd_1 + \rd_2 ) E_- ] \xi_- = 0 .
\label{bse}
\eea
As shown in Ref.\cite{shin}, from the form of the solutions (\ref{ansaa}),
we find that the electromagnetic current and the induced charge vanish;
\be
J^i_\pm = \bar\psi \gamma^i P_\pm \psi = 0, \qquad
G^0_\pm = \epsilon^{0ij} \rd_i J_{\pm, j} = 0 .
\label{vash}
\ee
By using Eqs.(\ref{eon}) and (\ref{vash}), the magnetic fields reduce to
\be
B_{\pm} = -F_{\pm, 12} = \pm {e \over \kappa} \rho_\pm
\label{mag}
\ee
where $\rho_\pm \equiv \psi^\dagger_\pm \psi_\pm$.
We note that Eqs.(\ref{eon}) can be written as
\be
\rd_j F^{ji}_\pm = \mp el \epsilon^{ij0} \rd_j J_{\pm, 0} .
\label{fieq}
\ee
From Eqs.(\ref{mag}) and (\ref{fieq}),
We can find the constant $l$ to be $l = - {1 \over \kappa}$.

To find the general solutions for the self-dual equation (\ref{ase}) and
(\ref{bse}), we note when $\eta_+$ and $\xi_-$ are decomposed into
its phase and amplitude,
\be
\eta_+ = \sqrt{\rho_+} e^{i\omega_+} , \qquad
\xi_- = \sqrt{\rho_-} e^{i\omega_-} ,
\label{dec}
\ee
the Eqs.(\ref{ase}) and (\ref{bse}) can be written as
\be
e \nabla \times \vec A_\pm = \nabla \times (\nabla \omega_{\pm} )
- \nabla \times (\nabla E_\pm ) t \mp {1 \over 2} \nabla^2 \ln \rho_{\pm} .
\label{bli}
\ee
From Eqs.(\ref{mag}) and (\ref{bli}) we obtain the equations for the
charge density $\rho_\pm$ :
\be
\nabla^2 \ln \rho_{\pm} = - {2e^2 \over \kappa} \rho_{\pm} .
\label{liu}
\ee
Eqs.({\ref{liu}) are the Liouville equations which are completely integrable.
When we take the solutions (\ref{ansaa}), $\kappa > 0$ is required in order
to have nonsingular positive solutions for the density $\rho_+$ and $\rho_-$.
The most general circularly symmetric nonsingular solutions to the
Liouville equations involve two positive constants $r_\pm$ and
${\cal N}_\pm$ \cite{solt} :
\be
\rho_\pm = {4 \kappa {\cal N}^2_\pm \over e^2 r^2}
           [ ( {r_\pm \over r })^{{\cal N}_\pm} +
           ( {r \over r_\pm })^{{\cal N}_\pm} ]^{-2}
\label{sola}
\ee
where $r_\pm$ are scale parameters.
In case of solutons (\ref{ansaa}), the magnetic dipole moments
are given by $\mu_\pm = \pm l Q_{\pm}
= \mp {Q_\pm \over \kappa} (\kappa > 0)$.

If we let
\be
\psi_+ = \eta_+ \left( \begin{array}{c} 1\\0 \end{array} \right) e^{-iE_+ t},
\qquad
\psi_- = \eta_- \left( \begin{array}{c} 1\\0 \end{array} \right) e^{-iE_- t},
\label{ansab}
\ee
the equations of motion (\ref{aeo}) and (\ref{beo}) become
\bea
&&(E_+ - m - el\epsilon^{ij0} \rd_j A_{+,i} ) \eta_+ = 0 ,\nonumber\\
&&D_{+ -} \eta_+ - it[(-i\rd_1 + \rd_2 ) E_+ ] \eta_+ = 0
\label{cse}
\eea
and
\bea
&&(E_- + m + el\epsilon^{ij0} \rd_j A_{-,i} ) \eta_- = 0 ,\nonumber\\
&&D_{- -} \eta_- - it[(-i\rd_1 + \rd_2 ) E_- ] \eta_- = 0 .
\label{dse}
\eea
If $\eta_\pm$ are decomposed into its phase and amplitude,
\be
\eta_+ = \sqrt{\rho_+} e^{i\theta_+} , \qquad
\eta_- = \sqrt{\rho_-} e^{i\theta_-} ,
\label{ded}
\ee
the amplitudes $\rho_\pm$ satisfy the Liouville equations
\be
\nabla^2 \ln \rho_{\pm} = \mp {2e^2 \over \kappa} \rho_{\pm}
\label{aliu}
\ee
where the $\mp$ signs are for the solutions of the forms $\psi_\pm$,
respectively.
Eqs.(\ref{aliu}) imply that there is no nonsingular positive solution
to satisfy the Liouville equations with respect to $\rho_\pm$ simultaneously.

If we let
\be
\psi_+ = \xi_+ \left( \begin{array}{c} 0\\1 \end{array} \right) e^{-iE_+ t},
\qquad
\psi_- = \eta_- \left( \begin{array}{c} 1\\0 \end{array} \right) e^{-iE_- t},
\label{ansac}
\ee
the equations of motion (\ref{aeo}) and (\ref{beo}) become
\bea
&&(E_+ + m - el\epsilon^{ij0} \rd_j A_{+,i} ) \xi_+ = 0 ,\nonumber\\
&&D_{+ +} \xi_+ - it[(i\rd_1 + \rd_2 ) E_+ ] \xi_+ = 0
\label{ese}
\eea
and
\bea
&&(E_- + m + el\epsilon^{ij0} \rd_j A_{-,i} ) \eta_- = 0 ,\nonumber\\
&&D_{- -} \eta_- - it[(-i\rd_1 + \rd_2 ) E_- ] \eta_- = 0 .
\label{fse}
\eea

From the form of the solutions (\ref{ansac}), the electromagnetic
current $J^i_\pm$, and the induced charge $G^0_\pm$ also vanish.
Therefore, the magnetic fields reduce to
\be
B_\pm = - F_{\pm , 12} = \pm {e \over \kappa} \rho_\pm
\label{maa}
\ee
and the constant $l$ to be $l = - {1 \over \kappa}$.

In this case,
we obtain the Liouville equations for the charge density $\rho_\pm$ :
\be
\nabla^2 \ln \rho_{\pm} = {2e^2 \over \kappa} \rho_{\pm} .
\label{lia}
\ee
As for Eqs.(\ref{lia}), $\kappa < 0$ is required in order to have
nonsingular positive charge density $\rho_\pm$.
In this case, the most general circularly symmetric nonsingular solutions
to the Liouville equations involve two positive constants $r_\pm$ and
${\cal N}_\pm$ \cite{solt} :
\be
\rho_\pm = - {4 \kappa {\cal N}^2_\pm \over e^2 r^2}
           [ ( {r_\pm \over r })^{{\cal N}_\pm} +
           ( {r \over r_\pm })^{{\cal N}_\pm} ]^{-2}
\label{solb}
\ee
In case of solutions (\ref{ansac}), the magnetic dipole moments
are given by $\mu_\pm = \mp {Q_\pm \over \kappa} (\kappa < 0)$.

If we let
\be
\psi_+ = \xi_+ \left( \begin{array}{c} 0\\1 \end{array} \right) e^{-iE_+ t},
\qquad
\psi_- = \xi_- \left( \begin{array}{c} 0\\1 \end{array} \right) e^{-iE_- t},
\label{ansad}
\ee
the equations of motion (\ref{aeo}) and (\ref{beo}) become
\bea
&&(E_+ + m - el\epsilon^{ij0} \rd_j A_{+,i} ) \xi_+ = 0 ,\nonumber\\
&&D_{+ +} \xi_+ - it[(i\rd_1 + \rd_2 ) E_+ ] \xi_+ = 0
\label{gse}
\eea
and
\bea
&&(E_- - m + el\epsilon^{ij0} \rd_j A_{-,i} ) \xi_- = 0 ,\nonumber\\
&&D_{- +} \eta_- - it[(i\rd_1 + \rd_2 ) E_- ] \xi_- = 0 .
\label{hse}
\eea
In this case,
we obtain the Liouville equations for the charge density $\rho_\pm$ :
\be
\nabla^2 \ln \rho_{\pm} = \pm {2e^2 \over \kappa} \rho_{\pm} .
\label{lib}
\ee
As for Eqs.(\ref{lib}), there is no nonsingular positive solution to
satisfy the Liouville equations.

In cases of solutions (\ref{ansaa}) and (\ref{ansac}),
the Hamiltonian density ${\cal H}$ reduces to
\be
{\cal H} = {1\over 2} F^2_{+, 12} + {1\over 2} F^2_{-, 12}
\pm m\rho_+ \pm m\rho_- + {e^2 l\over \kappa} \rho^2_+ + {e^2 l\over \kappa}
\rho^2_- ,
\label{ham}
\ee
respectively.
The total energy of the system can be written as
\bea
E =&& \int d^2 r {\cal H} = \int d^2 r [\pm m (\rho_+ + \rho_-)
-{e^2\over 2 \kappa^2} (\rho^2_+ + \rho^2_- )] \nonumber\\
&&= \pm m {2\kappa  \over e^2} ({\cal N}_+ + {\cal N}_- ) \nonumber\\
&&- {e^2 \over \kappa^2} [({4 \pi \kappa^2 {\cal N}^3_+
\Gamma (2- {1 \over {\cal N}_+ } ) \Gamma (2 + {1 \over {\cal N}_+ } )
\over 3e^4 r^2_+ })
+ ({4 \pi \kappa^2 {\cal N}^3_-
\Gamma (2- {1 \over {\cal N}_- } ) \Gamma (2 + {1 \over {\cal N}_- } )
\over 3e^4 r^2_- })] .
\label{enea}
\eea
As in Ref.\cite{shin}, $({\cal N}_+ , {\cal N}_-)$-soliton solutions
of this model (\ref{lagrang}) also have binding energies.
Thus the stability of the nontopological
$({\cal N}_+ , {\cal N}_-)$-solitons in this model is guaranteed by
the binding energies and the charge conservation laws of the system \cite{shin}.

From the Eqs.(\ref{ase}) and (\ref{bse}), we find
\be
E_+ = m - {e^2 \over \kappa^2} \rho_+ , \qquad
E_- = m - {e^2 \over \kappa^2} \rho_- .
\label{fena}
\ee
From the Eqs.(\ref{ese}) and (\ref{fse}), we also find
\be
E_+ = -m - {e^2 \over \kappa^2} \rho_+ , \qquad
E_- = -m - {e^2 \over \kappa^2} \rho_- .
\label{gena}
\ee
Therefore, the solutions (\ref{ansaa}) or (\ref{ansac})
is acceptable because the parity
transformation does not change any internal properties of the particle
like the energy except for the binding energy \cite{juns}.
However, from the Eqs.(\ref{cse}) and (\ref{dse}), we find
\be
E_+ = m - {e^2 \over \kappa^2} \rho_+ , \qquad
E_- = - m - {e^2 \over \kappa^2} \rho_- .
\label{fenb}
\ee
From Eqs.(\ref{gse}) and (\ref{hse}), we also find
\be
E_+ = - m - {e^2 \over \kappa^2} \rho_+ , \qquad
E_- = m - {e^2 \over \kappa^2} \rho_- .
\label{genb}
\ee

In these cases, we find that the definition of Eqs.(\ref{parity}) changes
the internal property under the parity transformation.
In other words, the parity transformation implies that
each solution transforms to the solution with the different energy.
This is unacceptable because the parity transformation should not
change any internal properties of the particles \cite{juns}.
Therefore, we can also ascertain that the forms (\ref{ansab}) and
(\ref{ansad}) are not the solutions of this model (\ref{lagrang}).

\begin{acknowledgments}
The author has benefited from conversations with J.M. Knight,
L.N. Chang and J.H. Yee and would like to thank Feng-Li Lin
for helpful discussion.

\end{acknowledgments}



\begin{references}

\bibitem{gern} 
S. Deser, R. Jackiw, and S. Templeton, Ann. Phys. (N.Y.) {\bf 140},
372 (1982); J. Hong, Y. Kim, and P.Y. Pac, Phys. Rev. Lett.
{\bf 64}, 2230 (1990); R. Jackiw and E.J. Weinberg, ibid. {\bf 64},
2234 (1990);
R. Jackiw, K. Lee and Erick J. Weinberg, Phys. Rev. {\bf D42},
3488 (1990); C. Lee, K. Lee and H. Min, Phys. Lett. {\bf B252}, 79 (1990).

\bibitem{ferm}
Y.M. Cho, J.W. Kim, and D.H. Park, Phys. Rev. {\bf D45}, 3802
(1992);
C. Duval, P.A. Horvathy, and L. Palla, ibid.
{\bf 52}, 4700 (1995).

\bibitem{shin}
S. Hyun, J. Shin, J.H. Yee and H. Lee, Phys. Rev. {\bf D55}, 3900
(1997).

\bibitem{solt}
R. Jackiw and S.Y. Pi, Phys. Rev. Lett. {\bf 64} 2969 (1990);
R. Jackiw and S.Y. Pi, Phys. Rev. {\bf D42} 3500 (1990).

\bibitem{juns}
J. Shin and J.H. Yee, Phys. Rev. {\bf D50}, 4223 (1994).

\bibitem{viol}
J. Fr\"ohlich and A. Zee, Nucl. Phys. {\bf B364}, 517 (1991);
K.B. Lyons, J. Kwo, J.F. Dillion, G.P. Espinosa, M. McGlashan-Powell,
A. Ramirez, and L. Schneemeyer, Phys. Rev. Lett. {\bf 64}, 2949 (1990);
S. Spielman, K. Fesler, C.B. Eom, T.H. Geballe, M.M. Fejer, and A. Kapitulnik,
Phys. Rev. Lett. {\bf 65}, 123 (1990);
H.J. Weber, D. Weitbrecht, D. Brach, A.L. Shelankov, H. Keiter, W. Weber,
Th. Wolf, J. Geerk, G. Linker, G. Roth, P.C. Splittgerber-H\"unnekes, and
G. G\"untherodt, Solid State Commun. {\bf76}, 511 (1990) .

\bibitem{inva}
J. Schonfeld, Nucl. Phys. {\bf B185}, 157 (1981) ;
G. Semenoff and N. Weiss, Phys. Lett. {\bf B250}, 117 (1990) ;
C.R. Hagen, Phys. Rev. Lett. {\bf 68}, 3821 (1992) .

\bibitem{topo}
D.G. Barci and L.E. Oxman, Phys. Rev. {\bf D52}, 1169 (1995);
M.E. Carrington and G. Kunstatter, ibid. {\bf 51}, 1903 (1995).

\bibitem{ycho}
Y.M. Cho, SNUTP 93-01 Preprint (1993)

\end{references}
\end{document}